\documentclass[twocolumn,10pt,prx,floatfix]{revtex4-1}

\input{glyphtounicode}
\usepackage[T1]{fontenc}
\usepackage[utf8]{inputenc}
\usepackage[english]{babel}
\usepackage{amsmath}  
\usepackage{amsfonts} 
\usepackage{graphicx} 
\usepackage[pdftex,bookmarks=true,bookmarksopen,bookmarksnumbered,
                colorlinks,
                linkcolor=blue,
                citecolor=blue]{hyperref}
\graphicspath{ {./} }
\usepackage{natbib}
\usepackage{array}
\usepackage[table]{xcolor}
\usepackage{ulem}
\usepackage{xcolor}
\usepackage{braket}
\usepackage{booktabs}
\usepackage{multirow}
\usepackage{caption}
\usepackage{float}

\newcommand{\aref}[1]{\hyperref[#1]{Appendix~\ref*{#1}}}

\begin{document}

\captionsetup[table]{name={TABLE},labelsep=period,justification=raggedright,font=small}
\captionsetup[figure]{name={FIG.},labelsep=period,justification=raggedright,font=small}
\renewcommand{\equationautorefname}{Eq.}
\renewcommand{\figureautorefname}{Fig.}
\renewcommand*{\sectionautorefname}{Sec.}

\title{The SMART protocol -- Pulse engineering of a global field for robust and universal quantum computation}

\author{Ingvild Hansen}
\author{Amanda E. Seedhouse}
\author{Andre Saraiva}
\author{Arne Laucht} 
\author{Andrew S. Dzurak}
\author{Chih Hwan Yang}
\affiliation{School of Electrical Engineering and Telecommunications, The University of New South Wales, Sydney, NSW 2052, Australia}
\date{\today}

\begin{abstract}

Global control strategies for arrays of qubits are a promising pathway to scalable quantum computing. A continuous-wave global field provides decoupling of the qubits from background noise. However, this approach is limited by variability in the parameters of individual qubits in the array. Here we show that by modulating a global field simultaneously applied to the entire array, we are able to encode qubits that are less sensitive to the statistical scatter in qubit resonance frequency and microwave amplitude fluctuations, which are problems expected in a large scale system. We name this approach the SMART (Sinusoidally Modulated, Always Rotating and Tailored) qubit protocol. We show that there exist optimal modulation conditions for qubits in a global field that robustly provide improved coherence times. We discuss in further detail the example of spins in silicon quantum dots, in which universal one- and two-qubit control is achieved electrically by controlling the spin-orbit coupling of individual qubits and the exchange coupling between spins in neighbouring dots. This work provides a high-fidelity qubit operation scheme in a global field, significantly improving the prospects for scalability of spin-based quantum computer architectures.
\end{abstract}


\maketitle

\section{Introduction}
\label{sec:intro}

Large-scale fault-tolerant quantum computing requires a robust and readily scalable qubit architecture, including initialisation, manipulation and measurement capabilities, with error rates below 1\,\% \cite{knill2005quantum,fowler2012surface}. This implies that high-performance qubit gates and, ultimately, long qubit coherence times are required. 
Several demonstrations of qubit fidelities $>99\,\%$ exist for small-scale qubit systems~\cite{muhonen2015quantifying,watson2018programmable,yoneda2018quantumdot,yang2019silicon,harty2014highfidelity,chow2009randomized,rong2015experimental,barends2014superconducting}, however a major obstacle on the way to realising a practical large-scale quantum computer is the challenge of scaling up architectures while maintaining high fidelities.

One strategy explored in the literature to overcome this problem is the electromagnetic dressing of qubits. By constantly driving the qubit, one can prolong the coherence times by continuously refocusing qubits against slow fluctuations in Rabi frequency~\cite{timoney2011quantum,laucht2017dressed,wu2019adiabatic,Cai2012LonglivedDS}. In addition, this is a scalable control scheme, since the driving microwave field can be applied globally to the entire multi-qubit device~\cite{kane1998siliconbased,veldhorst2017silicon,seedhouse2021quantum,vahapoglu2021singleelectron,jones2018logical}, so long as individual control of the Rabi frequencies to locally address qubits is possible. However, this type of global control is compromised by variability in qubit characteristics.

Spin qubits in silicon~\cite{zwanenburg2013silicon} are well suited to dressing, offer the prospect of individual addressability, and have excellent potential for large-scale integration due to their ability to leverage manufacturing from the microelectronics industry~\cite{veldhorst2017silicon,gonzalez-zalba2020scaling}. However, for silicon spin qubits, even in an isotopically purified substrate~\cite{itoh2014isotope}, residual nuclear spins \cite{hensen2020silicon} and spin-orbit-coupling \cite{veldhorst2015twoqubit,ruskov2018electron} due to interface disorder reduce both the coherence time and the homogeneity of the spin qubit properties.

Improvement in the robustness of dressed qubits can be achieved via the use of pulse engineering. Numerical algorithms like GRadient Ascent Pulse Engineering (GRAPE) \cite{khaneja2005optimal} have, in earlier work, been applied to construct optimal control pulses tackling such problems in order to improve gate performance \cite{yang2019silicon,zeng2019geometric}. A generalisation of the dressed qubit framework to the case of engineered electromagnetic pulses can be achieved by targeting specific types of qubit errors that are most commonly encountered across quantum computing platforms.

In this work we show that by combining microwave dressing with pulse shaping, that is, by modulating the amplitude of an always-on global field, we can realize a Sinusoidally Modulated, Always Rotating and Tailored (SMART) protocol for spin qubit operation that is readily scalable, with greater robustness to qubit variability and noise from microscopic sources, as well as noise from the control and measurement setup. We begin by briefly discussing dressed qubits in \autoref{sec:foreword}. The main principle of the SMART protocol is discussed in \autoref{sec:Tippe}, followed by the strategies for SMART qubit two-axes control in \autoref{sec:control}. The resulting one-qubit gate fidelities under a model of Gaussian noise are presented in \autoref{sec:fidelities}. We then discuss in further detail the implications of an always-on field for other aspects of universal quantum computing, taking as an example spins in silicon in \autoref{sec:fidelities2}. We focus on two-qubit gate fidelities in the presence of noise, as well as initialisation and readout. Finally, a summary of our conclusions regarding the feasibility of a quantum computer architecture employing this SMART protocol are presented in \autoref{sec:summary}.

\section{Foreword on dressed qubits}
\label{sec:foreword}

Most qubit systems are defined by a physical two-level system (such as a spin $1/2$, or two levels in an atom, for example) under static electromagnetic fields (either intrinsic to the qubit device or applied externally). Oscillatory electromagnetic fields are then applied in order to perform qubit control operations, which will serve as the tools for implementing logical gates. Alternatively, a qubit can be defined in terms of the dynamical states of the two-level system as driven by the externally applied oscillatory electromagnetic field. This is the case for a dressed qubit~\cite{mollow1969power,xu2007coherent,timoney2011quantum,laucht2017dressed}, which consists of a qubit that is permanently driven by an always-on resonant field. 

For a dressed qubit, $|0\rangle$ and $|1\rangle$ states are described in the laboratory frame as the qubit states that rotate with either the same phase or the opposite phase with relation to the driving field. Logical gates connecting the two states are then implemented by either speeding or delaying the precession of the qubit with regard to the driving field, changing the relative phase. The main advantage of encoding qubits in the driven state is that it provides dynamical decoupling from the environmental noise.  

Two elements limit the ability of the dressing scheme to refocus qubits under noise. Firstly, refocusing is only efficient if the time correlations of the noise amplitude exceed the Rabi period, which means that the spectral components of noise with frequency similar and above the Rabi frequency still impact the qubit coherence. Dynamical decoupling is unable to cope with this type of noise. 

The second limitation is noise causing large deviations in qubit Larmor frequency, which would cause the qubit to drift out of resonance with the microwave driving field and jeopardise the driving mechanism. This type of high-amplitude fluctuations usually occur in the form of a slow drift, such that for a few qubits this can be compensated by calibrating the microwave frequency between experiments. For multiple qubits this strategy of recalibration becomes inefficient. Moreover, applying different frequencies to each qubit would not allow for an unified global driving field, requiring individually focused driving fields which can be hard to implement in a full scale architecture.

In the dressed qubit strategy, the tolerance for deviations in resonance between the microwave and the qubit (or, equivalently, the tolerance for slow noise amplitudes) is set by the Rabi frequency. Pulse engineering~\cite{khaneja2005optimal,yang2019silicon,barnes2015robust}, however, can be used to develop improved driving strategies that have superior tolerances and are able to address noise in other parameters, such as fluctuations in the Rabi frequency.

\section{The SMART qubit protocol}
\label{sec:Tippe}

We introduce here a method of dressing the qubit with an oscillatory driving field that has a time-dependent amplitude, effectively creating a time-dependent Rabi frequency. Tailoring the amplitude modulation frequency to be in a certain proportion with the Rabi frequency, we are able to cancel different types of noise. The laboratory frame Hamiltonian of an arbitrary modulated driving field $\Omega(t)$ is given here by

\begin{equation}
    H_{\text{lab}} = \frac{h}{2}\left(\nu(t)\sigma_z+\Omega{(t)}2\cos{(2\pi{f}_{\text{mw}}t)}\sigma_x\right).
    \label{eq:gl0}
\end{equation}

In general, one can target multiple types of noise by adding different frequency and phase components to the amplitude modulation. We look at the special case where the global field amplitude is modulated by a single sinusoid, in which case the laboratory frame Hamiltonian is given by
\begin{equation}
    H_{\text{lab}}^{\text{sin}} = \frac{h}{2}\left(\nu(t)\sigma_z+\Omega_{\text{R}}\sqrt{2}\sin{(2\pi{f}_{\text{mod}}t)}2\cos{(2\pi{f}_{\text{mw}}t)}\sigma_x\right).
    \label{eq:gl2}
\end{equation}

\onecolumngrid

\begin{figure*}[h]
    \centering
    \vspace{-0.1cm}
    \includegraphics[width=\textwidth]{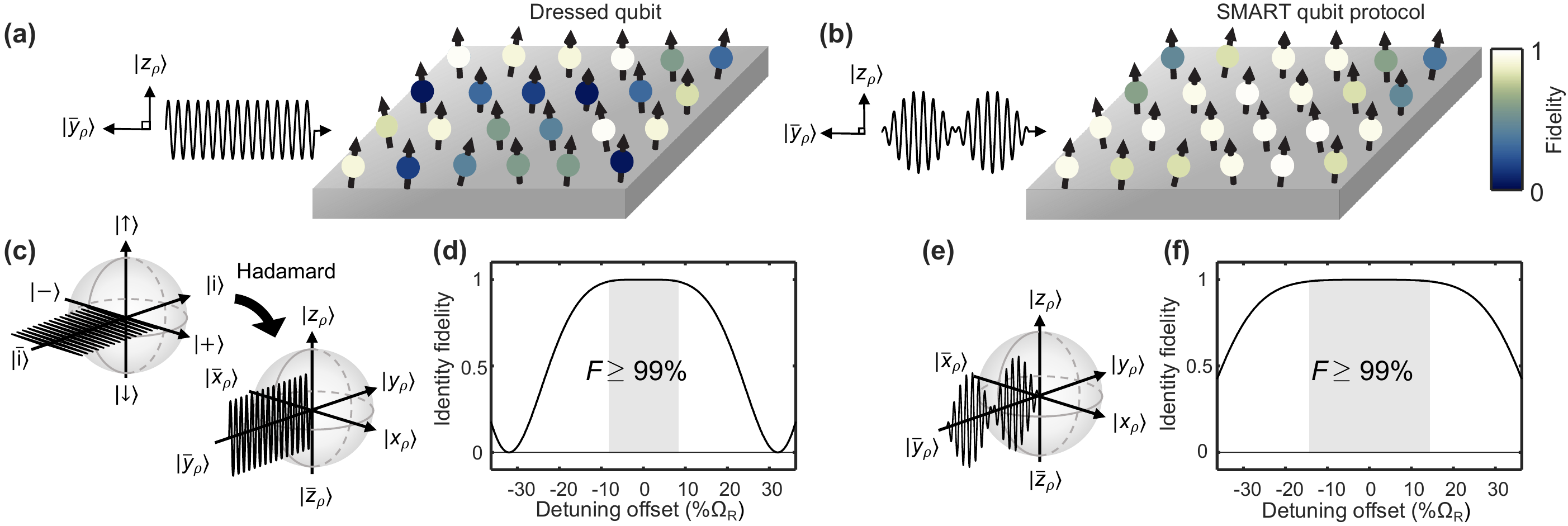}
    \caption{A qubit ensemble driven collectively by a global field consisting of (a) a continuous drive
    and (b) a sinusoidal modulated field. The Bloch spheres for the continuous drive is shown in c) together with the transformation from the rotating to the dressed spin frame. In d) the identity operator fidelity for a range of detuning offsets is shown where the range with fidelities above $99$\,$\%$ has been shaded. The Bloch sphere and the identity operator fidelities for the sinusoidal modulated case is given in e) and f). Here $\Omega_{\text{R}}=1$\,MHz.
    }
    \label{fig:gl}
\end{figure*}
\twocolumngrid

\noindent Here, $\sigma_x$ and $\sigma_z$ are Pauli matrices acting on the qubit state and $h$ is the Planck constant. The qubit Larmor frequency $\nu(t)$ has a time dependence that is controllable by external fields, and will be used to tune the resonance between the qubit and the driving field frequency ${f}_{\text{mw}}$ for controlled qubit rotations. The maximum amplitude of the oscillatory field creates Rabi rotations of frequency $\Omega_{\text{R}}\sqrt{2}$ on the qubit. This amplitude is modulated by the $\sin{(2\pi{f}_{\text{mod}}t)}$ term, where the modulation frequency ${f}_{\text{mod}}$ is a parameter that is chosen in order to optimise the noise cancelling properties of the driving field. The factor of $\sqrt{2}$ is a scaling factor in order to compare the resulting efficiency of the driving field in the cases of dressed and SMART qubits when adopting the same root mean square power of the global field. In \autoref{fig:gl} the fidelity of an identity operation on a dressed and SMART qubit ensemble is compared for different frequency detuning offset, showing higher robustness to resonance frequency variability for the latter.

The mathematical description and computer simulation of the qubit dynamics are significantly simplified when the Hamiltonian is written in the rotating frame that precesses with the same frequency as the driving field ${f}_{\text{mw}}$. In this case, 
\begin{equation}
    H_{\text{rot}}^{\text{sin}} = \frac{h}{2}\left(\Delta\nu(t)\sigma_z+\Omega_{\text{R}}\sqrt{2}\sin{(2\pi{f}_{\text{mod}}t)}\sigma_x\right),
    \label{eq:gl3}
\end{equation}
where $\Delta\nu(t)=\nu(t) - {f}_{\text{mw}}$ is the detuning between the controllable Rabi frequency and the driving field. 

Dressed qubit logical states are then the $|+\rangle$ and $|-\rangle$ states in the rotating frame, that is, the states parallel and anti-parallel to the $x$ axis in a rotating frame. We highlight this fact by referring to a \textit{dressed basis}, which is simply a Hadamard transformation over the rotating frame basis described in Eq.~\ref{eq:gl3}. This returns the logical qubit states to the conventional $z$ axis. Operating in the dressed basis implies the following axes transformations from the rotating frame basis: $|\hspace{-0.16cm}\uparrow\rangle\rightarrow|x_{\rho}\rangle$, $|\hspace{-0.16cm}\downarrow\rangle\rightarrow|\bar{x}_{\rho}\rangle$, $|+\rangle\rightarrow|z_{\rho}\rangle$, $|-\rangle\rightarrow|\bar{z}_{\rho}\rangle$, $|i\rangle\rightarrow|\bar{y}_{\rho}\rangle$ and $|\bar{i}\rangle\rightarrow|y_{\rho}\rangle$, hence a rotation about the $x$-axis in the dressed basis is equivalent to a rotation about the $z$-axis in the rotating basis etc. This change of quantisation axis can be seen from the Bloch sphere in \autoref{fig:gl}(c) where the qubit states along the conventional quantisation axis $|\hspace{-0.11cm}\uparrow\rangle$ and $|\hspace{-0.11cm}\downarrow\rangle$ now are in the equatorial plane.

The Hamiltonian in the dressed basis reads
\begin{equation}
    H_\rho^{\text{sin}} = \frac{h}{2}\left(\Omega_{\text{R}}\sqrt{2}\sin(2\pi{f}_{\text{mod}}t)\sigma_z+\Delta\nu(t)\sigma_x\right).
    \label{eq:gl}
\end{equation}

In general, the amplitude and frequency of the modulated global field determine its noise-cancelling properties. This example of a sinusoidal modulated global field can be extended to more sophisticated combinations of modulation components in order to cancel multiple types of noise as well.

To understand why a SMART qubit can be superior to a dressed qubit in terms of coherence time and gate performance, we derive an analytical expression for a model of quasi-static noise using the Magnus expansion series and analyse the noise cancelling properties using the geometric formalism from Ref.~\onlinecite{zeng2019geometric}. The geometric formalism is based on a description of the time evolution of the qubit $U(t)$ in terms of a three dimensional trajectory $\vec{s}(t)$ extracted from the first term in the Magnus expansion series. This trajectory is directly related to the microwave amplitude modulation $\Omega(t)$ through its curvature $\kappa$ ($\Omega(t)=\kappa(t)$). This is explained in more detail in \aref{app:magnus}. 

\begin{figure}[htb!]
    \centering
    \includegraphics[width = 0.42\textwidth]{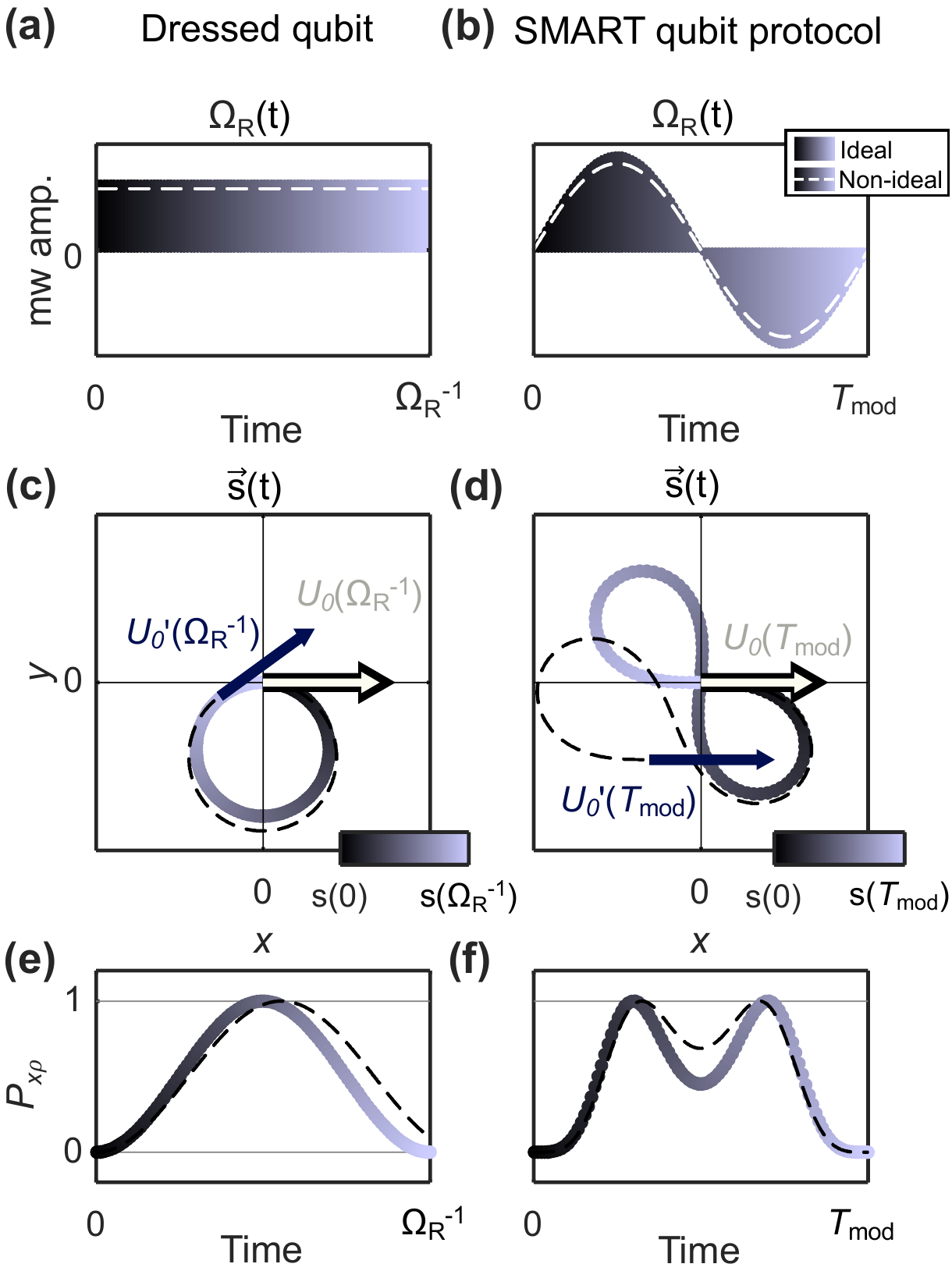}
    \caption{Geometric formalism describing noise-cancelling properties. (a-b) Global field amplitude modulation $\Omega(t)$ and (c-d) the corresponding space curve $\vec{s}(t)$, for the dressed and SMART qubit.
    The ideal modulation condition is plotted with a solid black line and a non-ideal condition with a dashed black line. The slope of the space curve at the end-point for the ideal (beige) and non-ideal case (blue) is plotted in (c-d) together with a black arrow representing the slope at the starting point. 
    The evolution of a qubit initialised to $|\bar{x}_{\rho}\rangle$ is shown in (e-f).}
    \label{fig:shapedqubit}
\end{figure}

In \autoref{fig:shapedqubit}(a-d) $\Omega(t)$ and $\vec{s}(t)$ are shown in the cases of dressed and SMART qubits. Both cases show a closed space curve (solid black line), a circle for the dressed case and a figure eight for the SMART case. This indicates cancellation of first order noise.  
The dashed lines in \autoref{fig:shapedqubit}(a,b) show examples where the amplitude is offset by some form of noise (such as fluctuation in source power) for the same gate time, resulting in a non-closed space curve or equivalently only partial first order noise cancellation. Information about the single-qubit gate can be found in the slope of $\vec{s}(T)$ at $t=0$ relative to $t=T$. Parallel slopes correspond to the identity operator and perpendicular slopes correspond to $\sqrt{\text{X}}$ and $\sqrt{\text{Y}}$ gates etc.

The geometric requirement for first order noise to cancel -- that $\vec{s}(t)$ is a closed curve ($\vec{s}(0)=\vec{s}(T)$) -- is achieved for the SMART qubit in \autoref{fig:shapedqubit}(b,d) by synchronising the modulation frequency $f_{\rm mod}$ in a certain proportion to the Rabi frequency $\Omega_{\text{R}}$. In order for second order noise to cancel as well, the area projected from the trajectory $\vec{s}(t)$ onto the $xy$-, $xz$- and $yz$-plane must all equal zero. The sign of a projected area is determined by the winding direction of the trajectory. Hence, the figure eight trajectory followed by the SMART qubit in \autoref{fig:shapedqubit}(d) has a positively signed lobe in the fourth quadrant and a negatively signed lobe in the second quadrant. The projected areas therefore sum to zero for the SMART qubits, but not for the dressed qubits. This higher order of noise cancellation translates into an improved tolerance to noise amplitudes in the case of SMART qubits, while the dressed qubit only provides first order noise cancellation.

Note that what we refer here as quasi-static noise could be originated in the stochastic electromagnetic fields of the quantum processor, but it may also have its origin in other non-idealities, such as crosstalk between qubits, fabrication variability and small unaccounted Hamiltonian terms (such as long-range dipolar coupling between spins or cross-Kerr interactions in superconducting qubits coupled through a bus cavity).

We mathematically find the optimal modulation conditions for the global field by forcing the first and second order Magnus expansion series terms to zero \cite{zeng2019geometric}. The optimal modulation frequency $f^{\text{opt}}_{\text{mod}}$ is found to have the following relation with $\Omega_{\text{R}}$
\begin{equation}
    f^{\text{opt}}_{\text{mod}}=\Omega_{\text{R}}\sqrt{2}/{j_i},
    \label{fopt}
\end{equation}
where $j_i$ is the $i$-th root of the Bessel function of zeroth order and $j_1=2.404826$. The derivation of this relationship can be found in \aref{app:magnus}. The duration of one period of the global field is denoted $T_{\text{mod}}$. For a SMART qubit initialised in the plane perpendicular to the global field axis, driven at $f^{\text{opt}}_{\text{mod}}$ with amplitude $\Omega_{\text{R}}\sqrt{2}$ and $\Delta\nu(t)=0$, a positive rotation of $\sim3\pi/2$ followed by a negative rotation of the same angle occur for every $T_{\text{mod}}$ of the global drive. The dressed qubit, on the other hand, continuously rotates without change in angular velocity. This is shown in \autoref{fig:shapedqubit}(e,f). The back-and-forth rocking of the SMART qubit and the continuous rotation of the dressed qubit about the global field axis both contribute to the continuous echoing of low frequency noise in these encoding strategies.

\section{SMART qubit two-axes control}
\label{sec:control}

Rotations using the SMART protocol in the dressed basis are achieved by applying frequency detuning $\Delta \nu(t)$ to the qubit with sinusoidal modulation at certain frequency and phase. The global field is always on, providing dynamically protected gates. 

Detuning of individual qubits can be implemented, for example, by pulsing the gate electrode above a spin qubit in semiconductors with spin-orbit coupling, effectively shifting the gyromagnetic ratio \cite{laucht2015electrically,kane1998siliconbased}; by modulating

\onecolumngrid

\begin{figure*}[h]
    \centering
    \vspace{0.2cm}
    \includegraphics[width=\textwidth]{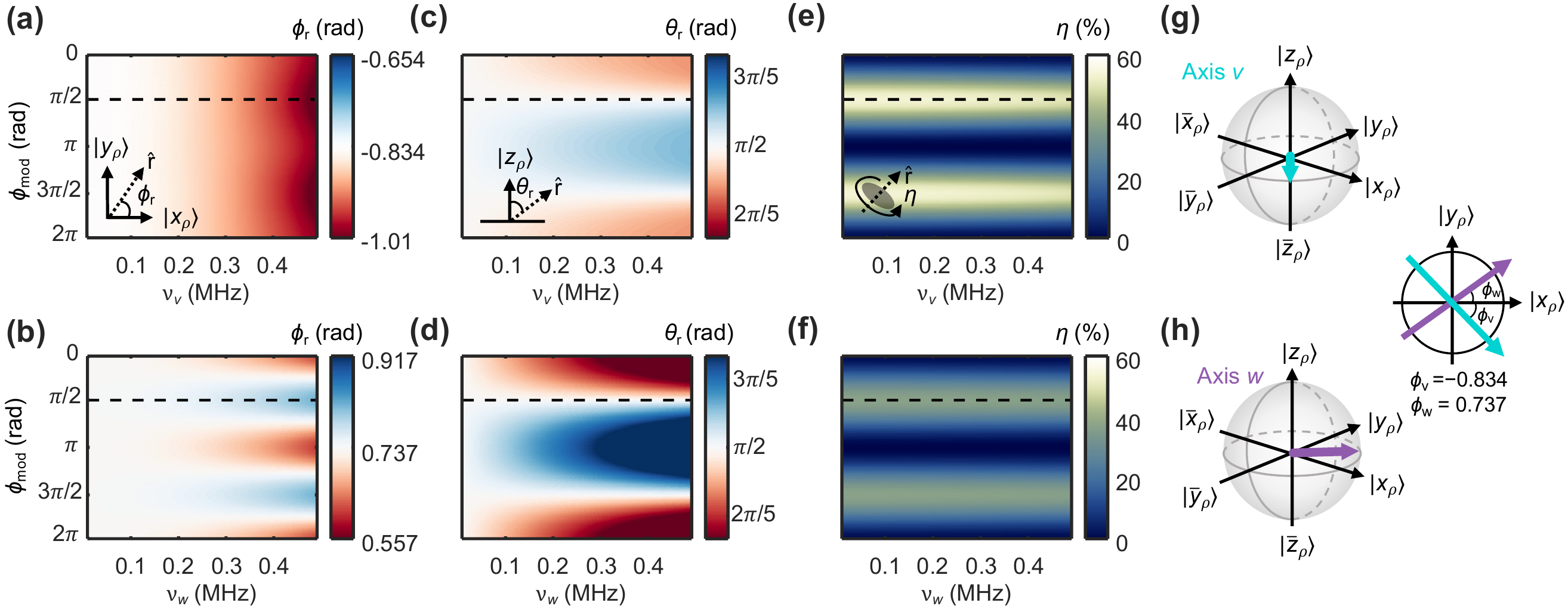}
    \caption{Rotation axis parameters (a) $\phi_{r}$ and (b) $ \theta_{r}$ for axis $v$ (top) and $w$ (bottom) of the SMART qubit for different values of the control amplitude $\nu_{v,w}$ and the phase offset between the microwave and the gate modulation $\phi_{\text{mod}}$. Phase $\pi/2$ has been indicated with a dashed horizontal line. The rotation efficiency $\eta$ calculated according to \autoref{eq:eta} is shown in (c) with the maximum values of 53.9\,\% and 37.3\,\% for axis $v$ and $w$, respectively. For small values of $\nu_{v,w}$ and $\phi_{\text{mod}}=\pi/2$ the resulting pair of perpendicular axes of rotation are illustrated on Bloch spheres in (g-h) with a relative angle of $\phi_{v}=0.834$ radians to the dressed $xy$ axis system. Here $\Omega_{\text{R}}=1$ MHz.
    }
    \label{fig:controlaxes}
\end{figure*}
\twocolumngrid

\noindent the hyperfine coupling between an electron and the static spin of the nucleus ~\cite{laucht2015electrically,thiele2014electrically,sigillito2017allelectric}; by locally changing the magnetic flux in a Josephson junction~\cite{krantz2019quantum}; and so on. 

Controlled rotations about one axis $v$ using sinusoidal local detuning of the qubit are described in the dressed basis by the Hamiltonian
\begin{equation}
\begin{split}
H_{\rho}^{v} &=H_{\text{global}}+H_{\text{local}}\\ &=H_{\text{global}}+\frac{h}{2}(\nu_{v}\sin(2\pi{f}_{\text{mod}}t+\phi_{\text{mod}})\sigma_x).
\end{split}
\label{eq:h1}
\end{equation}
The first term $H_{\text{global}}$ is the global field with sinusoidal modulations $\frac{h}{2}\Omega_{\text{R}}\sqrt{2}\sin(2\pi{f}_{\text{mod}}t)\sigma_z$ from \autoref{eq:gl}. Now we add a local control term $H_{\text{local}}$ that will be responsible for addressing an individual qubit by modulating its Larmor frequency, where $\phi_{\text{mod}}$ is the phase offset between the microwave and the qubit Larmor frequency modulation and $\nu_{v}$ the detuning amplitude of the local control term. Note that specifying that the modulation frequency of the local control field is equal to the one of the global field determines the direction for this rotation axis $v$, which in principle is not one of the Cartesian axes defined before. 

For two-axis control a second rotation axis $w$ can be found with a Hamiltonian of the same form as $H_{\rho}^{v}$ but with the detuning modulated at twice the frequency
\begin{equation}
\begin{split}
H_{\rho}^{w} =H_{\text{global}}+\frac{h}{2}\left(\nu_{w}\sin(4\pi{f}_{\text{mod}}t+\phi_{\text{mod}})\sigma_x\right).
\end{split}
\label{eq:h2}
\end{equation}
Any other combination of odd and even harmonics would also achieve two axes control, as long as the modulation remains synchronised with the global field echoing condition. However, higher harmonics exhibit lower rotation efficiency (see filter function formalism in \aref{app:magnus}). The direction of $w$ is, again, not correlated to the Cartesian directions in the general case.

The effective rotation is calculated from the time-evolution operator 
\begin{equation}
\begin{split}
        U_{r}(\chi) &= \cos\left(\frac{\chi}{2}\right)\mathrm{I}-i\sin\left(\frac{\chi}{2}\right)(r_x\sigma_x+r_y\sigma_y+r_z\sigma_z),
\end{split}
\end{equation}
using the identities $\theta_{r}=\arctan(r_y/r_x)$ and $\phi_{r}=\arctan(r_z(r_x^2+r_y^2)^{-1/2})$, where $\hat{r}=[r_x,r_y,r_z]$ is the unit rotation vector. The rotation angle $\chi$ can be calculated from the identity component of the operator, and $\hat{r}$ reconstructed when the identity component has been subtracted. In \autoref{fig:controlaxes} $\phi_{r}$, $\theta_{r}$ and the rotation efficiency $\eta$ is given as a function of $\phi_{\text{mod}}$ and $\nu_{v,w}$ for axes $v$ and $w$. The rotation efficiency is calculated for sinusoidal control terms according to
\begin{equation}
    \eta_{v,w}=\frac{P_{\text{out}}}{P_{\text{in}}}=100\,\%\times\bigg(\frac{\chi_{v,w}^2}{\big(2\pi T_{\text{mod}}\big)^2\big(\frac{\nu_{v,w}}{\sqrt{2}}\big)^2}\bigg),
    \label{eq:eta}
\end{equation}
where $P_{\text{out}}$ is given by the squared angular velocity and the sum represents the root mean square of the control sinusoids of amplitude $\nu_{v,w}$. The rotation efficiency is $100\%$ for square pulse control of an undressed qubit and $50\%$ for frequency modulation resonance control of a dressed qubit ~\cite{laucht2017dressed,seedhouse2021quantum}. This shows that both $v$ and $w$ rotations have comparable control strength to the dressed qubit.

By choosing appropriate values for $\nu_{v,w}$ and $\phi_{\text{mod}}$, the two axes $v$ and $w$ can be made perpendicular. These values correspond to $\nu_{v,w} \ll \Omega_{\text{R}}$ and $\phi_{\text{mod}} = \pi/2$, giving $\phi_{w}-\phi_{v}\approx\pi/2$ and $\theta_{v,w}\approx\pi/2$, as shown in \autoref{fig:controlaxes}(g,h). Hence, we have constructed two-axes control by tailoring the amplitude and phase of two sinusoidal driving fields of frequency $f_{\text{mod}}$ and $2f_{\text{mod}}$. By combining the two driving field in a weighted sum arbitrary two-axes control, including Cartesian $xy$-axes, can be engineered as discussed in the following paragraph.

From now on we will assume $\phi_{\text{mod}}=\pi/2$ and replace the sine from the local control term in \autoref{eq:h1} and \autoref{eq:h2} with a cosine. The condition on $\nu_{v,w}$ and $\phi_{\text{mod}}$ from \autoref{fig:controlaxes} only guarantees that the two rotation axes $v$ and $w$ are perpendicular, they do not coincide with $|x_{\rho}\rangle$ and $|y_{\rho}\rangle$ on the Bloch sphere in \autoref{fig:controlaxes}(g,h). Instead, they are rotated $-0.834$ radians ($-47.8$ degrees) in relation to the Cartesian axes. For the sake of completeness, we show that, in order to produce actual $x$- and $y$-rotations, the control terms of $v$ and $w$ can be combined in a linear fashion 
\begin{equation}
\begin{split}
        H^{x,y}_\rho =H_{\text{global}}+\frac{h}{2}\bigg(\nu_{v}^{x,y}\big(\cos(2\pi{f}_{\text{mod}}t)-1\big)+\\\nu_{w}^{x,y}\big(\cos(4\pi{f}_{\text{mod}}t)-1\big)\bigg)\sigma_x.
\end{split}
\label{eq:xy}
\end{equation}
Here, additional $-1$ terms are added in order to force the control amplitude to start and end at zero, which is advantageous for experimental reasons as the control fields are limited by a finite rise time and power. In order to find the optimal values for $\nu_{v}$ and $\nu_{w}$, GRAPE is applied \cite{khaneja2005optimal,yang2019silicon}.

\vspace{3cm}

\begin{table}
    \centering
     \setlength{\tabcolsep}{0pt}
        \caption{Coefficients used to construct $\sqrt{\rm{X}}$ and $\sqrt{\rm{Y}}$ gates with \autoref{eq:xy} for different duration and $\Omega_{\rm{R}}=1\,\text{MHz}$.
        }
        \setlength\extrarowheight{2pt}
        \begin{tabular}{cc|cc|c}
        \hline
        \hline
        \multicolumn{2}{c|}{$\sqrt{\rm{X}}$} & \multicolumn{2}{c|}{$\sqrt{\rm{Y}}$} &\multicolumn{1}{c}{}  \\
        \hline
        \multicolumn{1}{c}{$\nu^{x}_{v}$ (MHz) $\hspace{0.1cm}$} 
        &\multicolumn{1}{c|}{$\nu^{x}_{w}$ (MHz)$\hspace{0.3cm}$}   
        &\multicolumn{1}{c}{$\nu^{y}_{v}$ (MHz) $\hspace{0.1cm}$}  
        &\multicolumn{1}{c|}{$\nu^{y}_{w}$ (MHz) $\hspace{0.1cm}$}
        &\multicolumn{1}{c}{$\hspace{0.3cm}$ $t$ ($T_{\rm{mod}})$} \\
        \hline
        0.1515 & 0.3336  & -0.2154& 0.2224  & 1\\
        0.0893 & 0.1579  &-0.1056 &  0.1136 & 2\\
        0.0620 & 0.0921  &-0.0701 &  0.0760 & 3\\
        0.0271 & 0.0366  &-0.0300 & 0.0327  & 7\\
        0.0190 & 0.0254  &-0.0210 &  0.0229 & 10\\
        \hline
        \hline
    \end{tabular}
    \label{tab:tab}
\end{table}

\begin{figure}[htb!]
    \centering
    \hspace{-0.4cm}
    \includegraphics[width = 8.8cm]{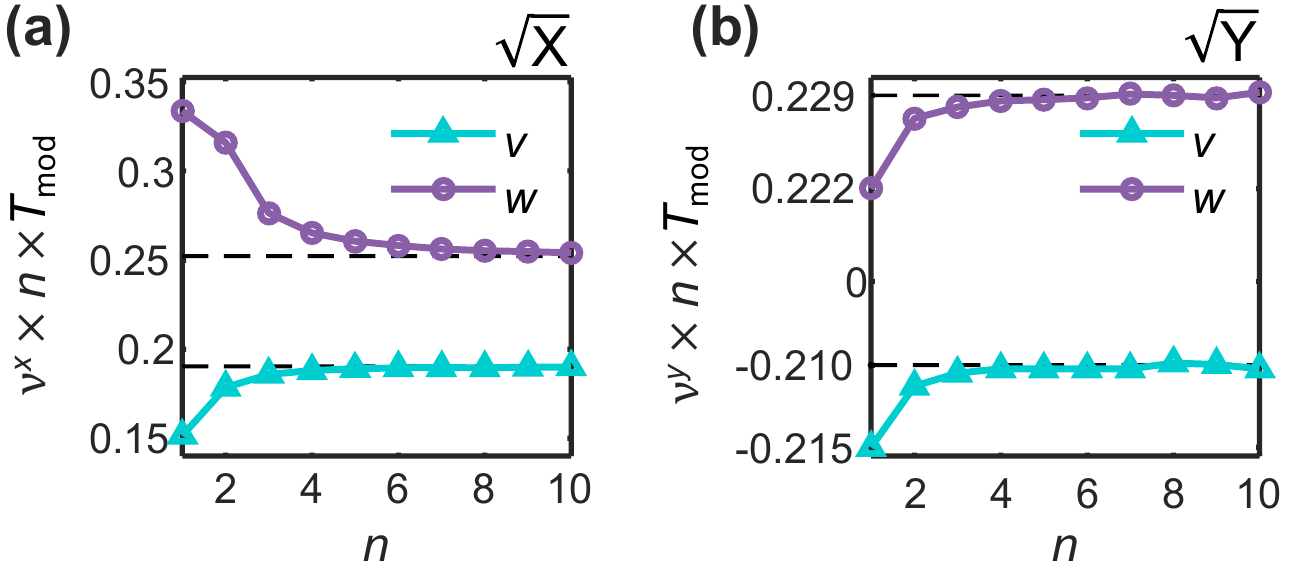}
    \caption{Coefficients $\nu_{v}$ and $\nu_{w}$ times the duration of a gate for (a) $\sqrt{\text{X}}$ and (b) $\sqrt{\text{Y}}$ gates for different gate duration according to \autoref{tab:tab}. The dashed horizontal line indicates the convergence value. Note that in (b) the $y$-axis is discontinuous.}
    \label{fig:mag}
\end{figure}

The duration of a one-qubit gate using the SMART protocol must equal a multiple $n$ of $T_{\text{mod}}$. For every $n$, the optimal values of $\nu_{v}$ and $\nu_{w}$ can be found from GRAPE. That is, each gate can be made to last for any integer number of $T_{\text{mod}}$. This is convenient as different systems can be limited by, for example, Larmor frequency tunability range or coherence times, in which case one would need longer or shorter gate duration, respectively. In \autoref{tab:tab}, values of $\nu_{v}$ and $\nu_{w}$ are given for a range of $n$. The same data multiplied by the gate duration is plotted in \autoref{fig:mag}(a,b), where the values clearly converge at longer gate duration. 
This convergence comes from the rotating wave approximation (RWA), where for large driving amplitudes (corresponding to short times in \autoref{fig:mag}) the approximation breaks down~\cite{laucht2016breaking}. There is a compromise between accurate rotation axes and fast control, as choosing a small integer number $n$ for the gate duration forces $\nu_{v}$ and $\nu_{w}$ to be higher in order to achieve the same rotation angle, affecting the accuracy of the rotation axis angles $\theta_{r}$ and $\phi_{r}$ found from \autoref{fig:controlaxes}. The fastest possible gate is limited by the amplitude of the Larmor frequency controllability in the system.

\begin{figure*}[h]
    \centering
    \vspace{0.4cm}
    \includegraphics[width = 1\textwidth]{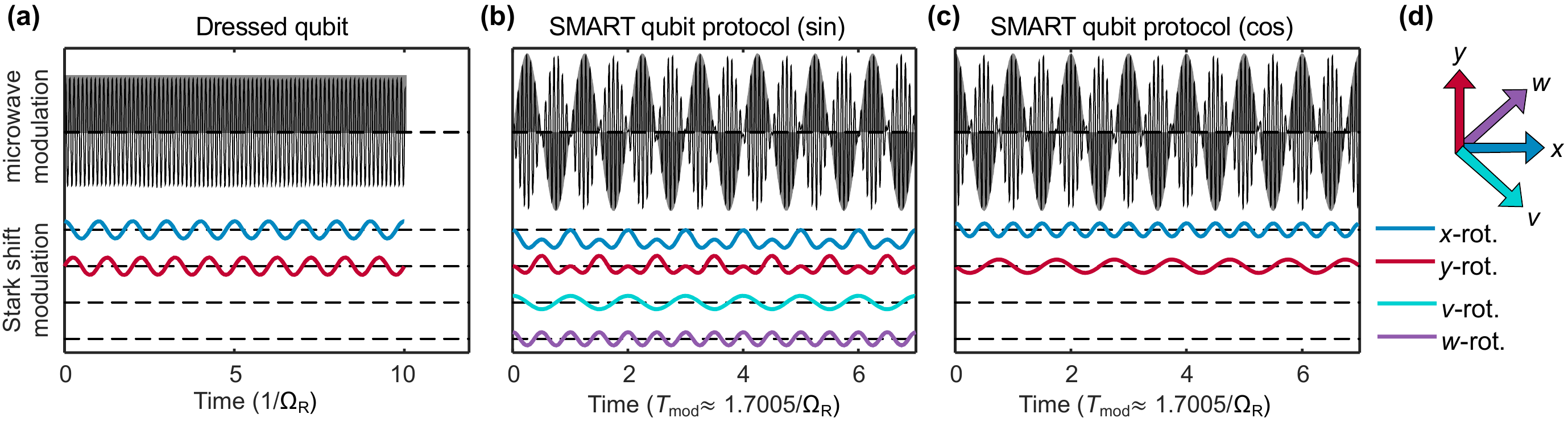}
    \caption{Global microwave field modulation and Stark shift control terms for (a) continuous drive $\sqrt{\text{X}}$ and $\sqrt{\text{Y}}$ gate and (b,c) SMART qubit $\sqrt{\text{V}}$, $\sqrt{\text{W}}$, $\sqrt{\text{X}}$ and $\sqrt{\text{Y}}$ gates. The gate durations are $10/\Omega_{\text{R}}$ for the dressed case and $7\times T_{\text{mod}}$ for the SMART case, and the relative microwave and Stark shift amplitude is to scale. In (d) the four axes are shown.}
    \label{fig:starkshiftcontrol}
\end{figure*}

It turns out that by modulating the global field with a cosine instead of a sine according to
\begin{equation}
    H_\rho^{\text{cos}} = \frac{h}{2}(\Omega_{\text{R}}\sqrt{2}\cos(2\pi{f}_{\text{mod}}t)\sigma_z+\Delta\nu(t)\sigma_x),
    \label{eq:gl22}
\end{equation}
$x$- and $y$-rotations can be achieved by simply single harmonic control terms without having to combine several harmonics in a linear fashion with coefficient extracted with GRAPE. However, the method developed here is useful for finding optimal parameters for arbitrary gate control strategies. The global microwave modulation and the Stark shift modulation for $x$-, $y$-, $v$- and $w$-rotation is shown in \autoref{fig:starkshiftcontrol} for the dressed and SMART qubit. 

\section{SMART qubit protocol gate fidelities}
\label{sec:fidelities}
In order to assess gate robustness to frequency detuning and microwave amplitude fluctuations using the SMART protocol, a noise analysis is carried out. Our noise model is a quasi-static Gaussian noise implemented in the system Hamiltonian as follows
\begin{equation}
        H_{\rho}=(1+\delta_{\Omega})\Omega_{\text{R}}\sqrt{2}\sin{(2\pi{f}_{\text{mod}}t)}\sigma_z+(\Delta\nu(t)+\delta_{\nu})\sigma_x.
    \label{eq:n}
\end{equation}
Here, $\delta_{\Omega}$ and $\delta_{\nu}$ represent the amplitude and detuning offset caused by the noise, respectively. The frequency detuning noise is considered as a simple offset, while the amplitude noise is taken to be proportional to the amplitude of the driving field. 

\begin{figure*}[h]
    \centering
    \includegraphics[width = \textwidth]{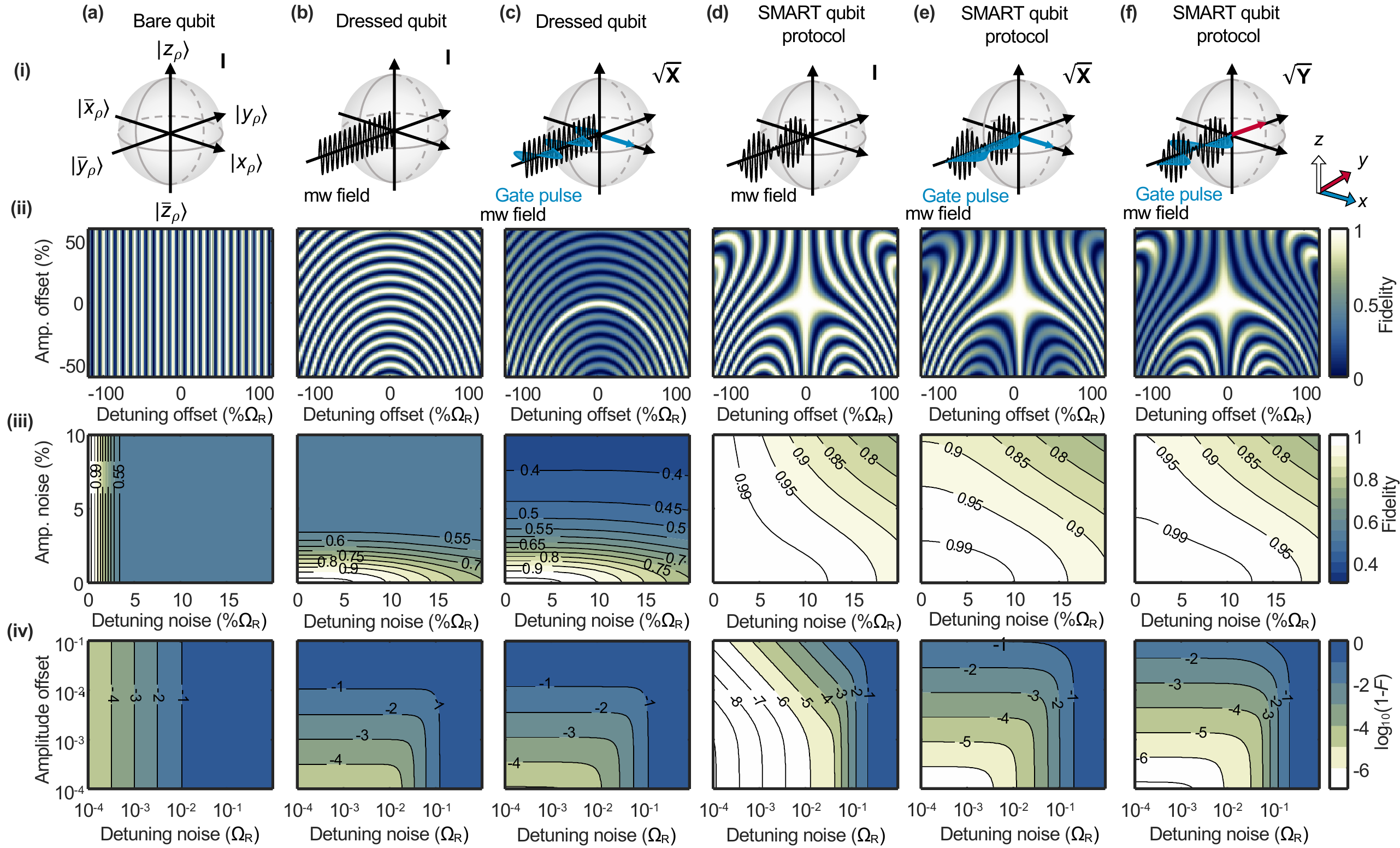}
    \caption{Gate fidelities for different values of amplitude and detuning offset/noise for the bare, dressed and SMART qubit. (a-c) show the identity gate fidelities for the bare and dressed qubit and a $\sqrt{\text{X}}$ gate for the dressed qubit, respectively. In (d-f) SMART qubit identity, $\sqrt{\text{X}}$ and $\sqrt{\text{Y}}$ is shown. Row (i) shows Bloch spheres with the relevant global field, local control field and rotation axis. In row (ii) the fidelity for offset values of amplitude and detuning is shown, and finally row (iii) and (iv) show Gaussian distributed noise in linear and log scale, respectively. Here $\Omega_{\text{R}}$ is $1$\,MHz and the gate durations are according to \autoref{fig:starkshiftcontrol}. The bare qubit identity gate has the same duration as the dressed gates.}
    \label{fig:comb}
\end{figure*}

In \autoref{fig:comb}(a-b) the fidelity of an identity gate is given for the bare (undressed) and the dressed qubit. A dressed $\sqrt{\text{X}}$ is shown in (c). The SMART qubit identity, $\sqrt{\text{X}}$ and $\sqrt{\text{Y}}$ gate is presented in (d-f). The first row shows fidelities corresponding to an operation generated by one fixed value of the offsets $\delta_{\Omega}$ and $\delta_{\nu}$, which represents one realisation of the noise. In the second and third rows, Gaussian averaging over several realisations has been applied, and it is shown in linear and logarithmic scales, 

\onecolumngrid

\twocolumngrid

\noindent respectively. The calculated fidelity for $\pi/2$ rotations are similarly given for (d) the dressed and (e,f) the SMART qubit protocol. More details on generating the 2D noise maps and 2D noise maps for $\sqrt{\text{V}}$ and $\sqrt{\text{W}}$ are provided in \aref{app:2dnoise} and \aref{app:vw}, respectively.

\section{SMART protocol for spin qubits}
\label{sec:fidelities2}

We now focus on the particular example of electron spin qubits in electrostatically confined quantum dots, in which the global driving can be performed through an oscillating magnetic field or, alternatively, an oscillating electric field that couples to spins through spin-orbit coupling. This spin-orbit coupling can also be used to control locally the value of the Rabi frequency through the Stark shift of the spin resonance frequency, which is a result of the influence of gate voltage bias on the effective $g$-factor of a spin in a given quantum dot.

All our numbers are chosen in the range of spin-orbit effects found in Si/SiO$_2$ electrostatic quantum dots, for which abundant literature exists to inform the expected variability and degree of controllability of the interface-induced spin-orbit coupling~\cite{jones2018logical,laucht2017dressed,veldhorst2015twoqubit,veldhorst2014addressable}.

For other qubit architectures the particular physical aspects of two-qubit gates, initialisation and readout may differ significantly and the feasibility of these operations under an always-on global field needs to be assessed case-by-case.

\begin{figure*}[htb!]
    \centering
    \includegraphics[width=1\textwidth,angle = 0]{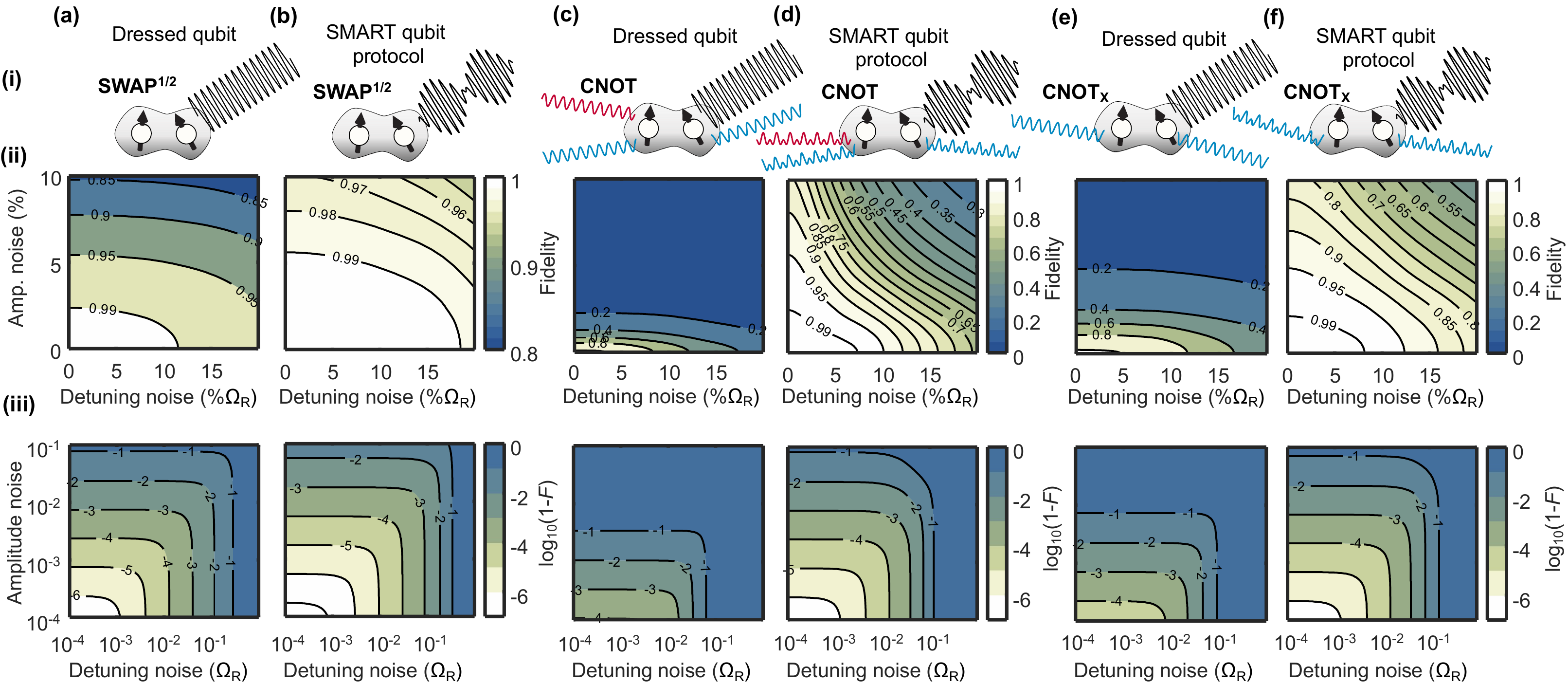}
    \caption{Two-qubit $\sqrt{\text{SWAP}}$ gate fidelities for different values of amplitude and detuning offset/noise for (a) the dressed qubit and (b) the SMART qubit. In (c) and (d) the gate fidelities for CNOT is given and in (e) and (f) for CNOT$_{\text{X}}$. Row (i) shows two qubits with a common global field and local Stark shift fields. In Row (ii) and (iii) the gate fidelities with Gaussian noise applied are shown on a linear and log scale, respectively. Here $\Omega_{\text{R}}$ is $1$\,MHz.
    }
    \label{fig:comb2}
\end{figure*}

\subsection{Two-qubit gates}

Two-qubit gates between spins based on exchange coupling can be implemented with a strategy similar to that of bare qubits. Applying voltage bias pulses to the electrostatic gates, the overlap between wavefunctions of neighbouring electrons can be tuned. In the example of bare qubits, the resulting spin-spin interaction depends on the ramp rates of the gate biases and the difference between qubit Larmor frequencies.

For the case of a driven qubit, such as the dressed or SMART qubit, the impact of the driving field on the resulting gate operation is also set by the exchange control ramp rates. The difference is that the relevant time scale is determined by the difference between Larmor and Rabi frequencies of the qubits. Further detail on the onset of the different operations for various ramp rates can be obtained in Ref.~\onlinecite{seedhouse2021quantum} in the case of dressed qubits.

Fidelity maps for the two-qubit gates $\sqrt{\text{SWAP}}$, $\text{CNOT}$ and $\text{CNOT}_{\text{X}}$ are given in \autoref{fig:comb2}. The $\sqrt{\text{SWAP}}$ gate is implemented assuming exchange gate control, where SWAP-like operation is the native two-qubit gate for qubits having the same resonance frequency \cite{seedhouse2021quantum}. The meaning of $\text{CNOT}_{\text{X}}$ here is a NOT operation on the target qubit conditional on the control qubit being $|x\rangle$ or $|\bar{x}\rangle$ instead of $|0\rangle$ or $|1\rangle$. The $\text{CNOT}$ and  $\text{CNOT}_{\text{X}}$ gate sequences used here are $(\sqrt{{\text{Y}}}^\dagger\otimes\text{I})\sqrt{\text{SWAP}}(\sqrt{{\text{X}}}^\dagger\otimes\sqrt{\text{X}})\sqrt{\text{SWAP}}(\sqrt{\text{Y}}\otimes\text{I})$ and $\sqrt{\text{SWAP}}(\sqrt{\text{X}}^\dagger\otimes\sqrt{\text{X}})\sqrt{\text{SWAP}}$. Here the assumption that the two qubits experience the same noise level is made (see \aref{app:2dnoise} for more details). For both one- and two-qubit gates the robustness to detuning and amplitude noise is seen to improve significantly compared to the bare and dressed case. 

\begin{figure*}[htb!]
    \centering
    \includegraphics[width =1\textwidth,angle = 0]{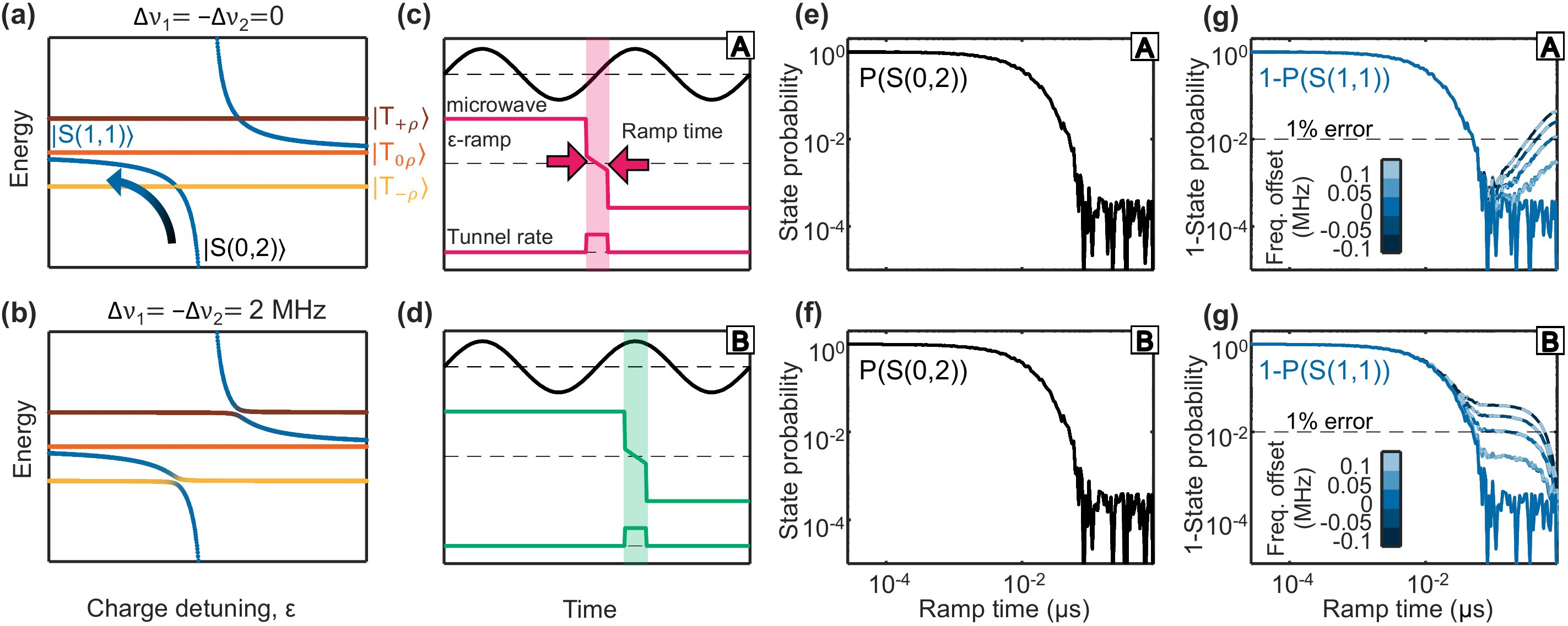}
    \caption{SMART two-qubit initialisation and readout. a) Energy diagram of the SMART two-qubit system for zero frequency detuning and for (b) $\Delta\nu_1=-\Delta\nu=0.2$\,MHz.  Initialisation of S(1,1) from S(0,2) with ramping centered about (c) the minimum microwave amplitude and (d) the maximum microwave amplitude of the global field. The results with different ramp rates and fixed charge detuning ramp range 50\,GHz $\rightarrow$ $-50$\,GHz ($\sim$0.2\,meV) is shown in (e-h) where the probability of S(0,2) and S(1,1) is plotted against ramp time. Fixed offsets in frequency detuning ($\Delta\nu_{1},\Delta\nu_{2}$) are introduced, with magnitudes given by the colorbar (two-colored dashed line representing the two qubits). Parameters used here include $\Omega_{\text{R}1}=\Omega_{\text{R}2}=1$\,MHz, $(\Delta\nu_1,\Delta\nu_2) \in\{0,\pm 0.05,\pm 0.1\}$\,MHz, $t_c=0.5$\,GHz. The total time is $2\times{T}_{\text{mod}}$.
    }
    \label{fig:init}
\end{figure*}

\subsection{Initialisation and readout}
\label{sec:initialisation}

High fidelity initialisation and readout are necessary for error-corrected quantum computing strategies. The constant driving field creates oscillations between the $|0\rangle$ and $|1\rangle$ states, which limit the range of strategies that can be used for initialisation and readout -- strategies based on energy-dependent transitions are hard to harmonise with a driving field. Both initialisation and readout are studied here in the context of strategies leveraging the Pauli spin blockade.

\begin{figure}[h!]
\hspace{-0.5cm}
    \centering
    \includegraphics[width =1\columnwidth,angle = 0]{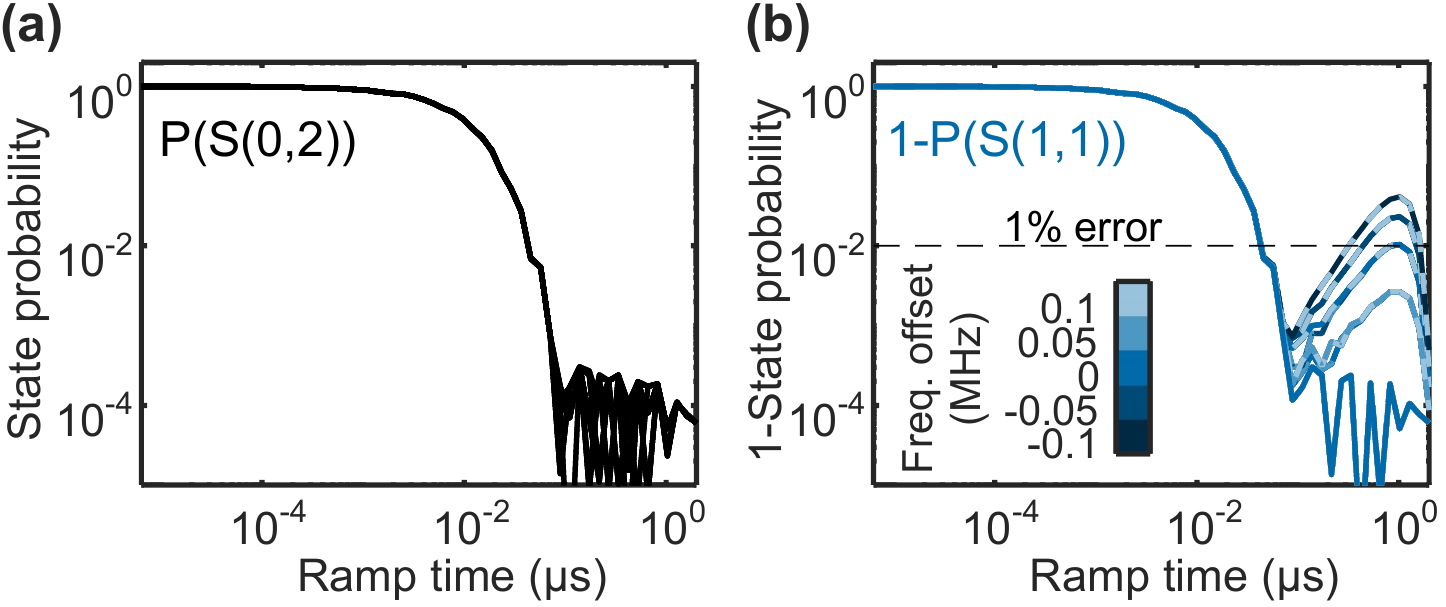}
    \caption{Dressed two-qubit initialisation for different ramp times and frequency detuning offsets. The state probability of S(0,2) and S(1,1) is shown with $\Omega_{\text{R}}=1$\,MHz and the total time $2/\Omega_{\text{R}}$.
    }
    \label{fig:init2}
\end{figure}

Initialisation of two-qubit SMART states is done similarly to the dressed spin qubit \cite{seedhouse2021quantum}, by ramping from negative to positive detuning at different rates. In \autoref{fig:init}(a) the energy diagram of the system as a function of charge detuning is shown at finite global field microwave amplitude with zero frequency detuning, and in (b) with $\Delta\nu_1=-\Delta\nu_2=0.2$\,MHz. For non-zero frequency detuning, an anticrossing appears and the ramping rate determines whether or not the spin crosses this energy gap diabatically. The system is initialised to an S(1,1) state from a S(0,2) state with a ramp centered about either (c) the minimum or (d) the maximum microwave amplitude ({\bfseries{A}} and {\bfseries{B}}). The transition from positive to negative detuning consists of a step before and after the slow ramp to achieve lower ramp rate, as seen from $\varepsilon$-ramp in \autoref{fig:init}(c,d). The state probability of S(0,2) and S(1,1) is given for different ramp times in \autoref{fig:init}(e-h) for the two cases. For comparison two-qubit dressed initialisation is shown in \autoref{fig:init2}. To show the robustness to resonance frequency variability different combinations of $\Delta\nu_1$ and $\Delta\nu_2$ $\in \{0,\pm0.05,\pm 0.1\}$\,MHz are simulated. A S(1,1) state is achieved with $>99$\,$\%$ fidelity after approximately 0.1\,\textmu s for case {\bfseries{A}} and 1\,\textmu s for case {\bfseries{B}} (at worst-case frequency offset). Centering the ramp about the minimum microwave amplitude ({\bfseries{A}}) looks to be a more robust options causing less mixing with the triplet states. This can be explained by looking at the effective echoing as a result of the global field after the ramp. For case {\bfseries{A}} close to a full period of the global field follows the ramp, whereas for case {\bfseries{B}} less than three quarters of a period. 

Readout can be performed similarly by reversing the process described above and relying on Pauli Spin blockade in the dressed frame \cite{seedhouse2021quantum}. In that case, the spin blockade is guaranteed for the duration of the spin relaxation time, and readout can be performed using some charge sensing technique. Note, however, that both the spin relaxation time and the charge readout bandwidth can be impacted by the field of the global driving, which can impose some engineering challenges. Further discussions on the engineering aspects of globally driven spin architecture are out of the scope of the present work, and some initial results in this direction can be seen in Refs.~\cite{vahapoglu2021singleelectron} and ~\cite{vahapoglu2021coherent}.

\section{Summary}
\label{sec:summary}

In this paper we propose to combine pulse engineering with electromagnetic dressing of qubits, in what we denote the SMART protocol. We have shown that qubits can be made more robust to detuning and amplitude variability caused by noise and/or sample inhomogeneity by applying a sinusoidal modulation to the global field, cancelling both first and second order noise terms in a Magnus expansion. By applying more complex modulation such as multi-tone driving, higher order noise terms can be cancelled as well. This is left for future work. We have analysed two-qubit gates, initialisation and readout in the particular example of spin qubits in quantum dots, however the SMART protocol can be applied to other systems with global coherent driving. The SMART protocol provides a clear and scalable path in terms of engineering constraints, making it a potential strategy for large-scale quantum computing.

\section*{Acknowledgments}

We acknowledge support from the Australian Research Council (FL190100167 and CE170100012), the US Army Research Office (W911NF-17-1-0198), and the NSW Node of the Australian National Fabrication Facility. The views and conclusions contained in this document are those of the authors and should not be interpreted as representing the official policies, either expressed or implied, of the Army Research Office or the U.S. Government. The U.S. Government is authorized to reproduce and distribute reprints for Government purposes notwithstanding any copyright notation herein. I.H and A.E.S acknowledge support from Sydney Quantum Academy.

\appendix

\section{Magnus expansion series}
\label{app:magnus}

In order to investigate how certain noise effects our system we look at the Hamiltonian in the interaction picture, found by transforming the noise Hamiltonian ($\delta\beta\sigma_i$) with the time evolution operator from the noiseless driving Hamiltonian $U(t)$. Here $\delta\beta$ is the fluctuation parameter on axis $i$ approximated as a constant for slow noise
\begin{equation}
    H_I(t)=U(t)^\dagger\sigma_i U(t)\delta\beta.
\end{equation}

The time evolution operator in the interaction picture is given by

\begin{equation}
    U_I(t)=\exp\left(\sum_{i=1}^{\infty}{A_i(t)}\right),
\end{equation}
and should equal identity for perfect noise cancellation. By truncating the sum we can find solutions where certain orders of noise cancels. For example, by making sure $A_1(T)$ is zero first order noise cancels out etc.

The first two order of the expansion of $U_I(t)$ include

\begin{equation}
    A_1(t)=\frac{1}{\delta\beta}\int_{0}^{t}{H_{\text{I}}(t_1)dt_1},
    \label{a1}
\end{equation}

\begin{equation}
    A_2(t)=\frac{1}{2\delta\beta^2}\int_{0}^{t}{dt_1}\int_{0}^{t_1}{dt_2[H_{\text{I}}(t_1),H_{\text{I}}(t_2)]}.
    \label{a2}
\end{equation}

The space curve parametrisation in \autoref{fig:shapedqubit}(b) is given by

\begin{equation}
    \vec{s}(t)=(x(t),y(t),z(t)),
\end{equation}
where $x(t),y(t),z(t)$ is extracted from $A_1(t)$ according to

\begin{equation}
    A_1(t)=x(t)\sigma_x+y(t)\sigma_y+z(t)\sigma_z.
\end{equation}

The deduction of \autoref{fopt} follows here corresponding to first and second order noise cancellation by forcing \autoref{a1} to zero. We can write \autoref{a1} in the form of a supermatrix $\hat{U}(t)$ using the identity $\text{vec}(ABC)=C^T\otimes{A}\hspace{0.1cm}\text{vec}(B)$ to allow for arbitrary noise axis

\begin{equation}
    \text{vec}(A_{1i}(t))=\frac{1}{\delta\beta}\int_{0}^{t}{U^T(t_1) \otimes U^\dagger(t_1)\delta\beta dt_1}\hspace{0.1cm}\text{vec}(\sigma_i).
\end{equation}

An arbitrary driving field $\Omega(t)$ about axis $\sigma_z$ (chosen to simplify maths, also dressed basis $x$-axis) gives us the time evolution operator

\begin{figure}[htb!]
    \centering
    \includegraphics[width=\columnwidth]{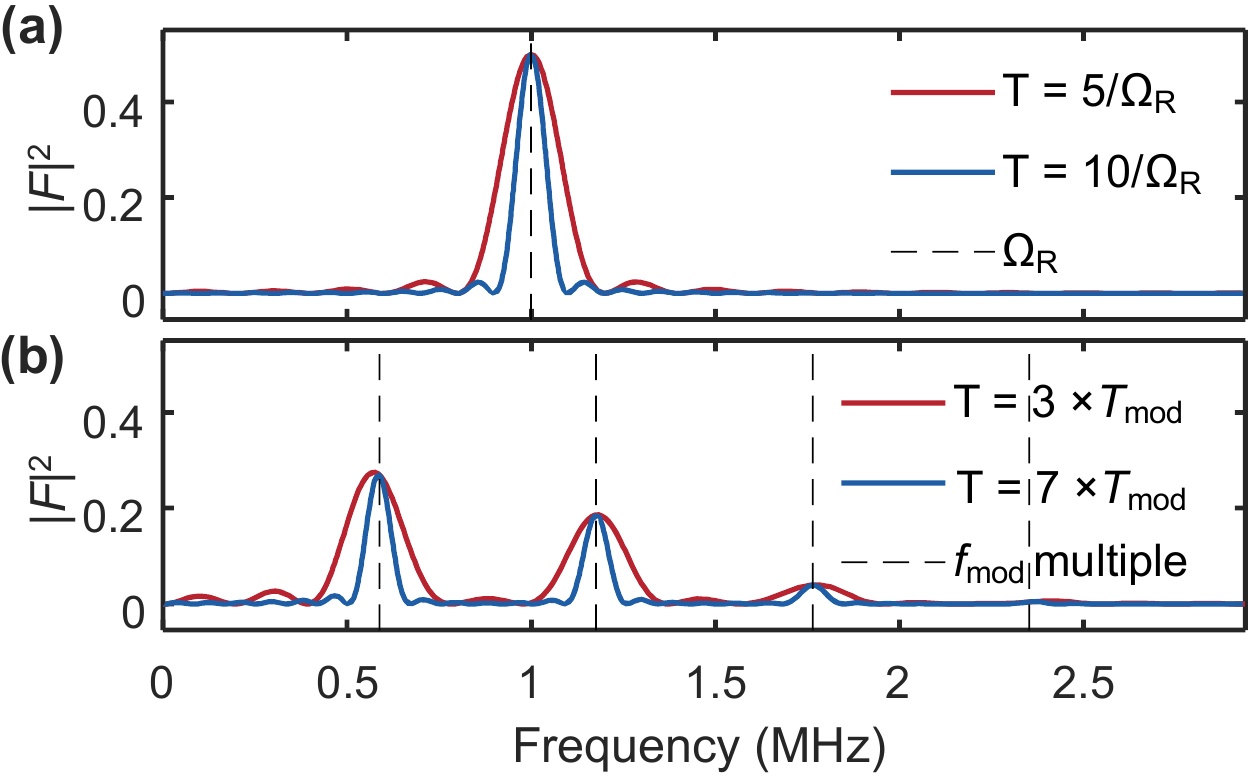}
    \caption{Filter function formalism applied to (a) the dressed and (b) the SMART qubit showing noise frequency susceptibility and equivalently controllability. Here $\Omega_{\text{R}}=1$\,MHz.
    }
    \label{fig:filter}
\end{figure}

\begin{equation}
    U(t)=\text{diag}(\exp
    \bigg\{-i\pi\int_{0}^{t}{\Omega(t)dt}\bigg\},\exp
    \bigg\{i\pi\int_{0}^{t}{\Omega(t)dt}\bigg\}),
\end{equation}
and the corresponding supermatrix
\begin{equation}
    \hat{U}(t)=\text{diag}(1,\exp
    \bigg\{-2i\pi\int_{0}^{t}{\Omega(t)dt}\bigg\},\exp
    \bigg\{2i\pi\int_{0}^{t}{\Omega(t)dt}\bigg\},1).
\end{equation}

Now looking at detuning noise in the dressed basis we find

\begin{equation}
    A_{1x}(t)=\text{mat}\bigg(\frac{1}{\delta\beta}\int_{0}^{t}{\Omega(t)\hat{U}(t)\delta\beta{dt} \hspace{0.1cm} \text{vec}(\sigma_x)}\bigg)
\end{equation}

For $A_{1x}(t)$ to be zero we have the following condition

\begin{equation}
    \int_{0}^{T}{\exp\bigg\{\pm2i\pi\int_{0}^{t}{\Omega(t_1)dt_1}\bigg\}dt}=0.
\end{equation}
Choosing a sine wave control with one period duration we get

\begin{equation}
        \int_{0}^{1/f}{\exp\bigg\{\pm 2i\pi\int_{0}^{t}{\Omega\sin(2\pi{ft_1})dt_1}\bigg\}dt}=0
\end{equation}

\begin{equation}
        \int_{0}^{1/f}{\exp\bigg\{\mp{i}\frac{\Omega}{f}\cos(2\pi{ft})\bigg\}dt}=0
\end{equation}

This can be recognised as one of Bessel's integrals with solution $\Omega/f=j_i$, where $j_i$ is the $i$-th root. It can be seen that $A_1$ is zero for $T=\frac{n}{2f}$. The second order term $A_2$ goes to zero when the time is chosen appropriately, which is $\frac{n}{f}$ in this case with any integer $n$. This is because the chosen control is a periodic and odd function, meaning that the second order cancellation happens for any values of $\Omega/f$ as long as the duration is a multiple of the period. This corresponds to $n$ loops through the space curve in \autoref{fig:shapedqubit}(d).

By substituting the fluctuation parameter with a tone of variable frequency ($\underline{\delta\beta}(f,t)=\delta\beta\exp(-2i\pi{f}t)$) we can probe the noise susceptibility of the SMART qubit at different frequencies (or similarly the controllability at certain control frequencies), according to filter function formalism. This is shown in \autoref{fig:filter}.

\begin{figure}[h]
    \centering
    \includegraphics[width=0.9\columnwidth,angle=0]{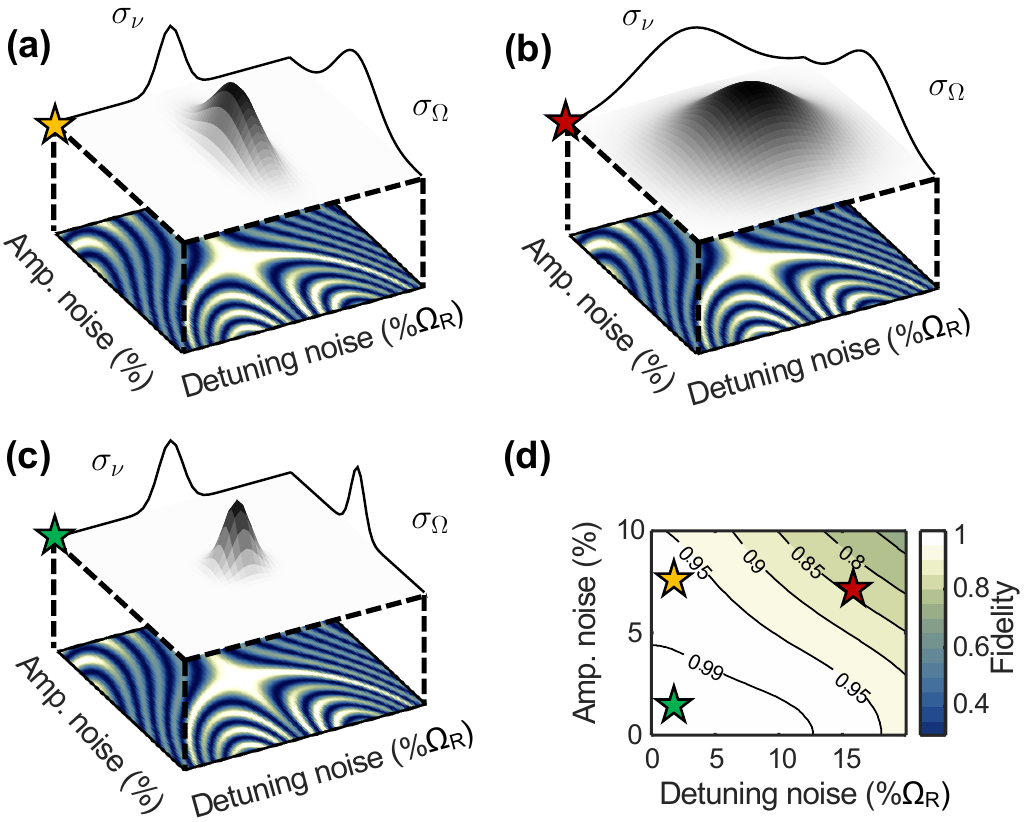}
    \caption{Schematic of method used to construct Gaussian noise model. The fixed offset noise map is multiplied by 2D Gaussians with $\sigma_{x}$ and $\sigma_{y}$ corresponding to the detuning and amplitude noise levels, shown here for three different cases (a-c). The stars in (d) indicate the following: low detuning noise and high amplitude noise (yellow), high detuning noise and high amplitude noise (red) and low amplitude noise and low detuning noise (green).}
    \label{fig:gn}
\end{figure}

\section{Simulation details of 2D noise maps}
\label{app:2dnoise}
For the 2D noise maps in \autoref{fig:comb} and \autoref{fig:comb2} the Hamiltonian given in \autoref{eq:n} is used. In order to generate the maps the following steps are executed:

\begin{enumerate}
    \item{Construct time-dependent Hamiltonian, $H$ with certain detuning and amplitude offset  $(\delta_{\nu},\delta_{\Omega})$ according to  \autoref{eq:n}.}
    \item{Time-evolve $H$ into time-evolution operator $U$ at a certain time.} 
    \item{Calculate the fidelity of the resulting operator $U$ by looking at the overlap with the target operator.}
    \item{Repeat the steps above for different amplitude and detuning offset values to create a 2D fidelity map $F(\delta_{\nu},\delta_{\Omega})$.}
    \item{Apply Gaussian averaging across the fixed noise map generated above, where the width of the applied Gaussian distribution is set by the noise level. That is, multiply the fixed noise map by a normalised 2D Gaussian around zero offset with widths ($\sigma_{\nu},\sigma_{\Omega}$) given by the noise levels (See \autoref{fig:gn}).
    }
\end{enumerate}

For the two-qubit case the noise levels of the two qubits are assumed to be the same. The same procedure as for the one-qubit gates is then followed, but integrating over all four noise dimensions $(\delta_{\nu{1}},\delta_{\nu{2}},\delta_{\Omega{1}},\delta_{\Omega{2}})$. Note that we assume there is no noise on the exchange coupling for the simulation.

\section{2D noise maps for axis \textit{v} and \textit{w}}
\label{app:vw}

\begin{figure}[htb!]
    \centering
    \includegraphics[width = 0.86\columnwidth,angle = 0]{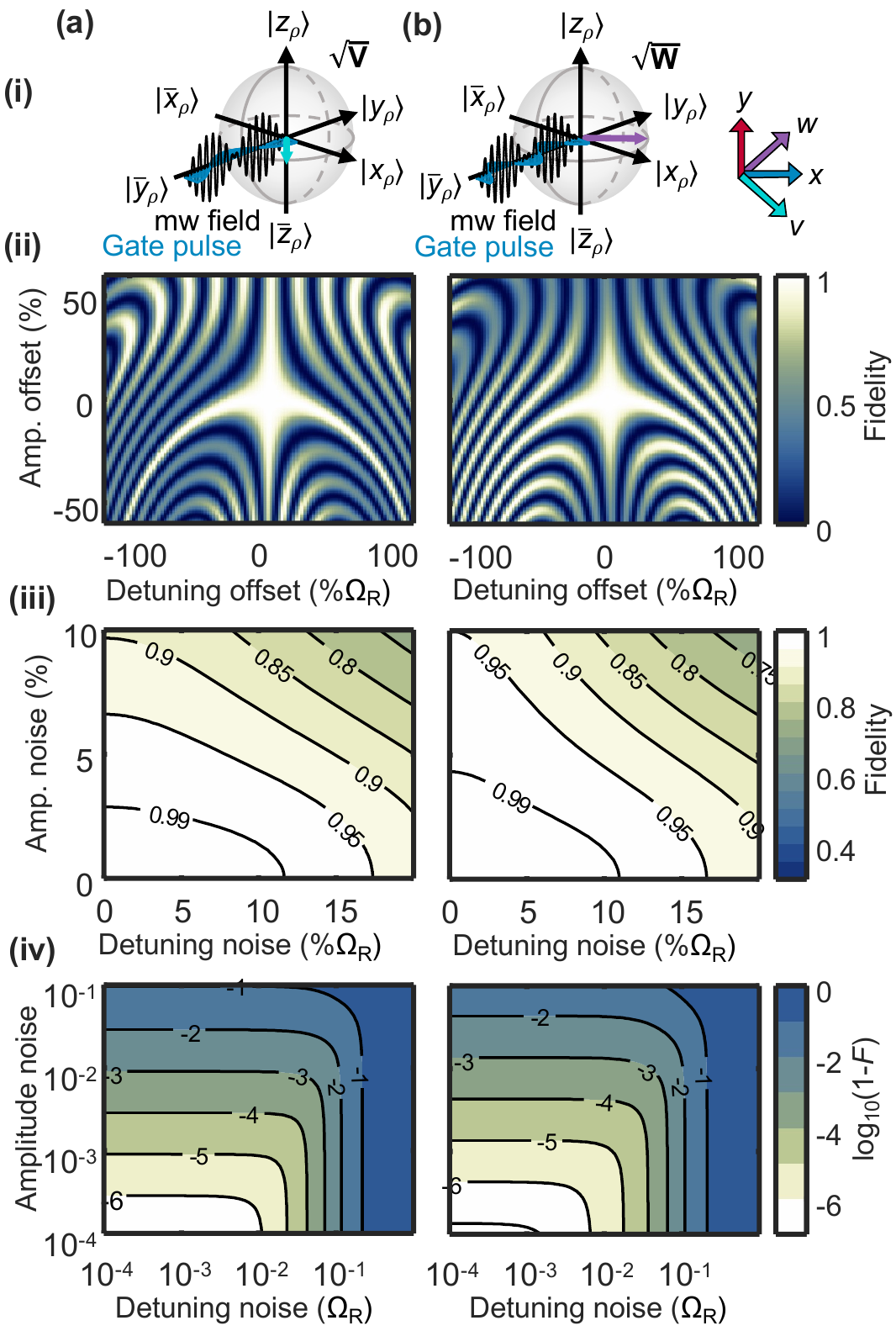}
    \caption{Gate fidelity maps for (a) the $\sqrt{{\text{V}}}$ gate and (b) the $\sqrt{{\text{W}}}$ gate. Row (i) shows Bloch spheres with the relevant global field, local control field and rotation axis. In row (ii) the fidelity for offset values of amplitude and detuning is shown, and finally row (iii) and (iv) show Gaussian distributed noise in linear and log scale, respectively.}
    \label{fig:vw}
\end{figure}

The 2D noise maps of $\pi/2$ rotation about the ${v}$ and ${w}$ axes in \autoref{fig:vw} are found to be similar to the SMART $\sqrt{\text{X}}$ and $\sqrt{\text{Y}}$ gates.

\bibliography{main}

\end{document}